\documentclass[12pt]{JHEP3}
\usepackage{epsfig}
\usepackage{amsmath}
\newcommand{\p}{\partial}
\newcommand{\Tr}{{\rm Tr}}
\title{
Recombination of Intersecting D-branes 
by Local Tachyon
 Condensation} 

\author{
Koji Hashimoto and Satoshi Nagaoka\\ 
Institute of Physics, University of Tokyo\\
Komaba, Tokyo 153-8902, Japan\\
E-mail: \email{koji@hep1.c.u-tokyo.ac.jp}, 
\email{nagaoka@hep1.c.u-tokyo.ac.jp}}

\abstract{We provide a simple low energy description of recombination of
intersecting D-branes using super Yang-Mills theory. The recombination
is realized by condensation of an off-diagonal 
tachyonic fluctuation localized at the intersecting point.
The recombination process is equivalent to brane-antibrane annihilation,
thus our result confirms Sen's conjecture on tachyon condensation, 
although we work in the super Yang-Mills theory whose energy scale is
much lower than $\alpha'$. We also discuss the decay width of
non-parallelly separated D-branes.}
 

\preprint{UT-Komaba/03-05, hep-th/0303204}

\begin{document}

\section{Introduction}

The significance of D-brane dynamics in string theories has been widely
recognized, and for unified formulation of nonperturbative string/M 
theory it is believed to be promising to adopt D-branes as fundamental
objects. Low dimensional D-branes, such as D-particles and D-instantons,
have been utilized along this direction \cite{matrix}, 
while higher dimensional
D-branes have relative difficulty due to the complicated 
interaction of the 
extended worldvolumes. One of such interaction is the 
{\it recombination}, or equivalently {\it reconnection}, 
of intersecting D-branes (especially D-strings). For fundamental
strings, the recombination effect has been introduced as an essential
interaction in an open string field theory \cite{HIKKO}.  
It strongly suggests that the recombination is one of the most important
mechanisms to be clarified, for formulating any field theory of
interacting D-branes. 
On the other hand, recombination of D-branes turned out to be an
intriguing realization of Higgs phenomena in Standard Model on
intersecting branes \cite{higgs}
and inflation in braneworld scenario \cite{cosmology}. 
The recombination appears as an essential dynamics 
in these recent developments in string phenomenology, hence its
precise formulation in string theory is required.

It is well-known \cite{pol}
that the recombination is due to the tachyon mode
appearing as the lowest mode of a string connecting the intersecting
D-branes (see Fig.~\ref{intersectfig}), although the mechanism 
has not been shown explicitly so far. As seen below, 
this is related to
Sen's conjectures \cite{Senconje}
stating that tachyon condensation on brane-antibrane is
equivalent to the disappearance of them.
When the intersection angle $\theta$ is $\pi$, the intersecting D-branes
are just a parallel brane-antibrane pair. The relevant tachyon mode
appears in excitation of a string connecting the two. If we decrease
$\theta$ a little, then this tachyonic mode is localized near the
intersection point. Provided the tachyon condensation makes the branes
disappear, it affects them only locally and the branes are recombined
(See Fig.~\ref{tachyonfig}). The verification of the Sen's conjectures,
especially on  
the disappearance of open string degrees of freedom, has not been shown
in a complete form yet.
So it would be possible to say that showing the D-brane recombination
explicitly is qualitatively equivalent to the verification of 
the Sen's conjectures. In addition,
the system of intersecting D-branes includes the parallel brane-antibrane
system as a special limit $\theta=\pi$, and naively it is expected that 
any multi-body systems of D-branes are involved with intersecting branes
generically, rather than its special case $\theta=\pi$ of the 
parallel brane-antibrane. This leads us to motivate the study of
the local tachyon condensation on the intersecting branes, which is the
recombination process.

In this paper, we provide a very simple description of the recombination
of intersecting D-branes by analysing corresponding background in 
super Yang-Mills theories which are low energy effective theories of
D-branes. We identify one 
of the fluctuation eigen modes with the tachyon state concerning the
annihilation of branes, and show that the condensation of this mode
exactly gives the recombination of the intersecting branes.
Our description provides a geometrical realization of the open string
tachyon condensation.

In Ref.~\cite{AkiWati}, fluctuation spectra of Yang-Mills theory
have been studied for
a T-dualized configuration of the intersecting branes wound on a torus. 
The authors of
Ref.~\cite{AkiWati} showed the agreement between the Yang-Mills 
fluctuation spectra and string theory spectra, while due to the
T-duality geometrical picture of the recombination was 
difficult to be seen. Our work does not involve the T-duality and fills
the gap.  

In Section 2 we study a 1+1 dimensional 
$SU(2)$ super Yang-Mills theory which is induced on parallel D-strings
at low energy.\footnote{Higher dimensional D-branes 
intersecting with a single angular parameter can be treated with a
trivial generalization of our work. }
We turn on a linear Higgs background in the diagonal component,
which amounts to turning on the intersecting angle $\theta$.
The fluctuation analysis around this background shows 
that there appears a tachyonic mode whose
mass squared (and its dependence on $\theta$) 
coincides with the worldsheet result of
Ref.~\cite{BDL}, for small $\theta$. 
After diagonalizing the fluctuation, 
we see that the tachyonic mode is exactly associated with the brane
recombination. 
All the computations are carried out in the low energy approximation
of string theory whose energy scale is much lower than the string scale
$\alpha'$. This is possible since at small $\theta$ the tachyon 
becomes almost massless.

We also attempt to analyse the behavior of $\theta \sim \pi$,
in Section 3. In the limit $\theta = \pi$, 
the brane configuration becomes a parallel coincident D1-antiD1 pair. 
We first clarify the relation between the D-brane recombination and 
the Sen's conjectures on the tachyon condensation on brane-antibrane. 
With natural ansatz from string scattering amplitudes, we adopt 
a tachyon action and its coupling to the gauge fields
living on the D-branes. This natural 
tachyon action is used to show that the 
worldsheet spectrum near $\theta=\pi$ is precisely reproduced in the
fluctuation analysis. Furthermore, the geometrical 
recombination is shown also in
this system $\theta  \sim \pi$ 
by analysing the backreaction to the transverse scalar field. 

In Section 4, we describe dynamical aspects of the recombination by 
evaluating the decay width of non-parallelly separated D-branes.
We show that for smaller intersection angle $\theta$, 
the decay probability gets smaller and the system approaches 
a supersymmetric system of parallel D-branes.
We speculate on how the recombination process proceeds. 

The low energy Yang-Mills approximation in Section 2
is valid for small intersection
angle $|\theta| \ll 1$, so in Appendix A
we include the non-Abelian Born-Infeld (NBI) corrections to it, to see
how the string spectrum with the higher order in $\theta$
is reproduced in the fluctuation spectrum. 
We use $(\alpha')^2F^4$ corrections in the NBI \cite{tse} and 
follow the techniques developed in Ref.~\cite{AkiWati},  
to find complete agreement in the spectra to the order $\theta^3$.

%%%%%%%%%%%%%%%%%%%%%%%%%%%%%%%%%%%%%%%%%%%%%%%%%%%%%%%%%%%
%%%%%%%%%%%%%%%%%%%%%%%%%%%%%%%%%%%%%%%%%%%%%%%%%%%%%%%%%%%
%%%%%%%%%%%%%%%%%%%%%%%%%%%%%%%%%%%%%%%%%%%%%%%%%%%%%%%%%%%

\section{D-brane recombination realized in Yang-Mills} 

%\subsection{Expected string spectrum}

\begin{figure}[bhtp]
\begin{center}
\begin{minipage}{13cm}
\begin{center}
\includegraphics[width=9cm]{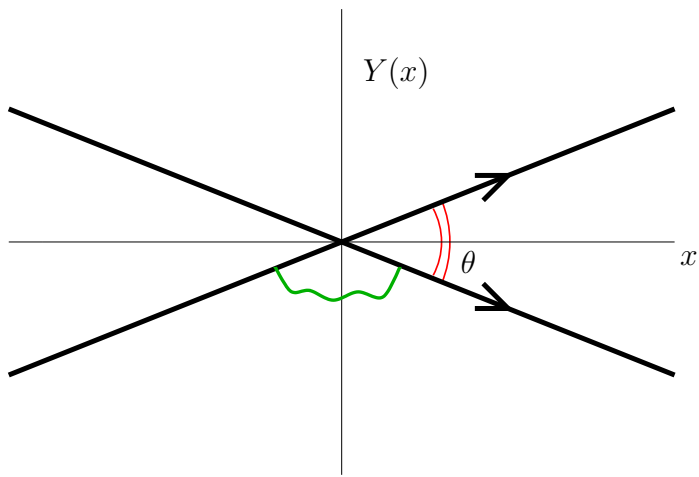}
\put(-120,150){$Y(x)$}
\put(0,80){$x$}
\put(-83,77){$\theta$}
\caption{Two D-strings are intersecting with each other at an angle
 $\theta$. The wavy line depicts a fundamental string
 connecting the D-strings. 
The worldvolumes are parametrized by the horizontal
 direction $x$ while its displacement from the $x$ axis is given by the
 transverse Higgs field $Y(x)$. }
\label{intersectfig}
\end{center}
\end{minipage}
\end{center}
\end{figure}

Parallelly placed multiple D-branes in type II superstring theories have
super Yang-Mills theory as their low energy description
\cite{Witten}. The non-Abelian structure stems from excitations of
strings connecting different D-branes (see Fig.~\ref{intersectfig}). 
When an intersection
angle $\theta$ ($0 < \theta < \pi$) is turned on, 
these strings get confined to the intersection point 
so that the energy coming from the tension times its length is
minimized. A part of the
spectrum of the Neveu-Schwarz sector of this localized open strings
was studied in Ref.~\cite{BDL, Jab} giving 
\begin{eqnarray}
 m^2 = \left(n-\frac12\right)\frac{\theta}{\pi\alpha'} \ ,
\label{1}
\end{eqnarray}
where $n$ $(\geq 0)$ is an integer, and $\theta$ is the intersection
angle ($\theta=0$ corresponds to a coincident parallel D-branes, 
see Fig.~\ref{intersectfig}).\footnote{Ref.~\cite{BDL} studied also the
Yang-Mills viewpoint.} In
this paper we treat the simplest case of turning on only a single
intersection angle, while generically with higher dimensional D-branes 
we may consider more parameters for the angles.

The important observation of Eq.~(\ref{1}) is that for $n=0$ the mass
squared becomes negative, and this is the tachyonic mode of our
concern. Since naively the intersecting branes can be recombined and
lower the total energy by shortening their lengths, it has been expected
that the tachyonic mode $n=0$ is an indication of this recombination
\cite{pol}. To 
show this mechanism explicitly is our aim of this paper.

%%%%%%%%%%%%%%%%%%%%%%%%%%%%%%%%%%%%%%%%%%%%%%%%%%%%%%%%

\subsection{Fluctuation analysis of Yang-Mills}

We start with a 1+1 dimensional $SU(2)$ Yang-Mills action
with maximal supersymmetries,\footnote{In this paper except Section
4, we omit writing
tensions of D-branes which appear as an overall
factor of the actions.} 
\begin{eqnarray} \label{YMH}
S= - \Tr 
\int\!dtdx^1\left[ \frac{1}{2}F_{\mu\nu}^2 
+ (D_\mu \Phi_i)^2 \right] \ .
\end{eqnarray}
In this expression we omit fermions. The indices $\mu,\nu=0,1$ are 
worldvolume coordinates while $i=2,3,\cdots,9$ is for transverse 
directions in the bulk. This action describes the low energy dynamics of
parallel two D-strings. 
The fields are decomposed into components by Pauli matrices as 
$A_\mu = (1/2)\sigma^a A_\mu^a$ and a similar expression for 
$\Phi_i$, with the definition of the covariant derivative as 
\begin{eqnarray}
F_{\mu\nu}^a \equiv \p_\mu A_{\nu}^a - \p_\nu A_\mu^a
+ \epsilon^{abc} A_\mu ^b A_\nu^c \ , \quad 
D_\mu \Phi^a \equiv \p_\mu \Phi^a + \epsilon^{abc} 
A_\mu^b \Phi^c \ .
\end{eqnarray}
The locations of the D-strings $Y^i$ are specified by the eigenvalues 
of the Higgs field with the rescaling $2\pi\alpha'\Phi_i = Y^i$. In
the following we turn on only $Y^9$ and omit the transverse index $i$. 

The background representing the intersecting D-strings should be 
\begin{eqnarray}
 \Phi^3 = q x \ , \quad A_\mu = 0 \ .
\label{backg}
\end{eqnarray}
Here $x$ denotes $x^1$. It is obvious that this is a solution of the
equations of motion for the action (\ref{YMH}). This solution breaks
all the supersymmetries. 
Taking into account the
rescaling factor $2\pi\alpha'$, one finds that 
the constant slope $q$ is related to the intersection angle $\theta$ as 
\begin{eqnarray}
 q = \frac{1}{\pi\alpha'}\tan(\theta/2) \ .
\label{relationq}
\end{eqnarray}
The brane configuration is shown in Fig.~\ref{intersectfig}.

Let us proceed to the analysis of the fluctuations around this
background (\ref{backg}).  
We turn on only $A_1^2\equiv a$ and $\Phi^1\equiv \varphi$ since
other fluctuation fields decouple from these,\footnote{
Another pair $(A_1^1,\Phi^2)$ which is decoupled from the above appears
with a similar Lagrangian. This another pair provides doubly degenerated
spectrum, which is consistent with the string spectrum that is also
doubly degenerated due to the possible orientations of the string. 
We shall see that in each sector a tachyonic mode appears which is
combined into a complex tachyon field.} 
at the quadratic
level. 
The Lagrangian quadratic in the fluctuations is calculated as 
\begin{eqnarray}
 L = \frac12 (\p_0a)^2 +\frac12 (\p_0\varphi)^2-\frac12
\left(
(\p_x\varphi)^2 + 2qxa\p_x\varphi + q^2 x^2 a^2 -2qa\varphi
\right) \ .
\label{fluctu}
\end{eqnarray}
The equations of motion for the fluctuations are
\begin{eqnarray}
 \left(
\begin{array}{cc}
-q^2x^2 -\p_0^2 & -qx\p_x + q \\
qx\p_x + 2q & \p_x^2 - \p_0^2
\end{array}
\right)
\left(
\begin{array}{c}
a(t,x) \\ \varphi(t,x)
\end{array}
\right)=0 \ .
\end{eqnarray}
To obtain the mass spectrum, we decompose the fluctuation fields
into the linear combination of mass eigen functions as
\begin{eqnarray}
\left(
\begin{array}{c}
a(t,x) \\ \varphi(t,x)
\end{array}
\right)=\sum_{n\geq 0}
\left(
\begin{array}{c}
\widetilde{a}_n(x) \\ \widetilde{\varphi}_n(x)
\end{array}
\right) C_n(t)
\end{eqnarray}
where the decomposed field $C_n(t)$ satisfies a free field equation
\begin{eqnarray}
(\p_0^2 + m_n^2) C_n(t)=0 
\label{free}
\end{eqnarray}
with the mass eigenvalues $m_n$
for each $n$. The eigen equations are given by 
\begin{eqnarray}
 \left(
\begin{array}{cc}
-q^2x^2 + m_n^2 & -qx\p_x + q \\
qx\p_x + 2q & \p_x^2 + m_n^2
\end{array}
\right)
\left(
\begin{array}{c}
\widetilde{a}_n(x) \\ \widetilde{\varphi}_n(x)
\end{array}
\right)=0 \ .
\end{eqnarray}
Solving 
this matrix differential equation, the mass squared is obtained as 
\begin{eqnarray}
 m_n^2 = (2n-1)q 
\label{specq}
\end{eqnarray}
for normalizable eigen functions 
$(\widetilde{a}_n(x),\widetilde{\varphi}_n(x))$,
with $n=0,2,3,4,5,\cdots$.  
The explicit expression for the eigen functions are
\begin{eqnarray} \label{sol1}
&&\widetilde{a}_n 
(x)=- e^{-qx^2/2} \sum_{j=0,2,\cdots}^n (-1)^{\frac{j}{2}} 
\frac{4^{\frac{j}{2}}}{j!}
\frac{n(n-2)\cdots (n-j+2)}{2n-1} (j-1) 
\left( x\sqrt{q/2}\right)^j \ , \nonumber\\
&&\widetilde{\varphi}_n 
(x)= e^{-qx^2/2}\sum_{j=0,2,\cdots}^n (-1)^{\frac{j}{2}} 
\frac{4^{\frac{j}{2}}}{j!}
\frac{n(n-2)\cdots (n-j+2)}{2n-1} (2n\!-\!j\!-\!1) 
\left(x \sqrt{q/2} \right)^j 
\nonumber
\end{eqnarray}
for $n=0,2,\cdots$, and
\begin{eqnarray}\label{sol2}
&&\widetilde{a}_n (x)=
- e^{-qx^2/2}\sum_{j=1,3,\cdots}^n (-1)^{\frac{(j-1)}{2}}
 \frac{4^{\frac{(j-1)}{2}}}{j!}
\left(\frac{j-1}{2}\right) (n-3) \cdots (n-j+2)
\left(x \sqrt{q/2} \right)^j \ , \nonumber\\
&&\widetilde{\varphi}_n (x)=
 e^{-qx^2/2} \!\left(\!\!\!\sqrt{q/2} x+\!\!\!\!\!\!
\sum_{j=3,5,\cdots}^n\!\!\!\! %(-1)^{\frac{(j-1)}{2}} 
\frac{(-4)^{\frac{(j-1)}{2}}}{j!}
\left(n\!-\!\frac{j\!+\!1}{2}\right) (n\!-\!3) \cdots (n\!-\!j\!+\!2)
\left(x \sqrt{q/2}\! \right)^j  
\right) \ \
\nonumber
\end{eqnarray}
for $n=3,5,\cdots$. 
Substituting the relation (\ref{relationq}) into the mass formula
(\ref{specq}), we find that it 
agrees with the string worldsheet spectrum 
(\ref{1}) up to ${\cal O}(\theta^3)$ ambiguity. In order to match two
results to higher order in $\theta$, we should have to adopt the 
non-Abelian Born-Infeld Lagrangian as a starting point. In appendix A, we
include the first nontrivial $\alpha'$ corrections of order 
$(\alpha')^2 F^4$ from string theory, and showed that the fluctuation 
spectrum agrees with the worldsheet result up to ${\cal O}(\theta^5)$
ambiguity. 

Our concern is the lowest mode $n=0$ which is tachyonic. Surprisingly,
the eigen functions for this lowest mode is quite simple, 
\begin{eqnarray}
 \widetilde{\varphi}_0(x) = \widetilde{a}_0(x) =  \exp 
\left[-\frac{qx^2}{2}\right]
\left(=\exp\left[
-\frac{\tan(\theta/2)}{2\pi\alpha'}x^2
\right]\right) \ .
\label{gauss}
\end{eqnarray}
The fluctuation modes are localized at $x=0$ which is the intersection
point, which agrees again with the worldsheet perspective.
If $\theta$ is small enough, the above fluctuation mode does not
invalidate the low energy approximation of the 
D-brane action as Yang-Mills theory, since the derivatives (involved
with the string $\alpha'$ corrections) acting on the fluctuation 
give only small contributions. In this respect, 
the approximation is invalid for $\theta$ 
exceeding $\pi/2$.

The fact that the eigen functions are Gaussian is consistent with the
T-dual results in Ref.~\cite{AkiWati} where the eigen functions are
$\Theta$ functions. The $\Theta$ functions are roughly a summation of
periodically placed Gaussian functions, and Ref.~\cite{AkiWati}
considered D-branes wrapping a torus for the T-duality so the above
localized eigenfunctions should appear periodically there.

\subsection{Realization of D-string recombination}

Let us find the constraint on the magnitude $C\equiv C_0(t)$ of the 
tachyonic fluctuation (\ref{gauss}) first. This is derived by requiring that
the quadratic approximation is valid, that is, the terms quartic in
fluctuations are much smaller than quadratic terms. Straightforward 
calculation shows that in the original Yang-Mills Lagrangian 
the quadratic terms with eigen functions
(\ref{gauss}) are 
\begin{eqnarray}
\text{(quadratic terms)} \sim qC^2 e^{-qx^2} \ ,
\end{eqnarray}
while the quartic terms are
\begin{eqnarray}
\text{(quartic terms)}\sim C^4 e^{-2 qx^2} \ .
\end{eqnarray}
Hence the above consistency requires 
\begin{eqnarray} \label{cond}
q \gg C^2 \ .
\label{constraint}
\end{eqnarray}
This is the condition for that the approximation we adopted for the
fluctuation analysis is valid. The free field
equation (\ref{free}) for $C(t)$ is solved with an exponentially growing
function  
\begin{eqnarray}
 C \sim e^{\sqrt{q}t} \ ,
\end{eqnarray}
and our approximation is valid until this $C(t)$ reaches the critical
value $\sqrt{q}$ as seen in Eq.~(\ref{constraint}). 

%\FIGURE[htp]{
\begin{figure}[tp]
\begin{center}
\begin{minipage}{12cm}
\begin{center}
\includegraphics[width=10cm]{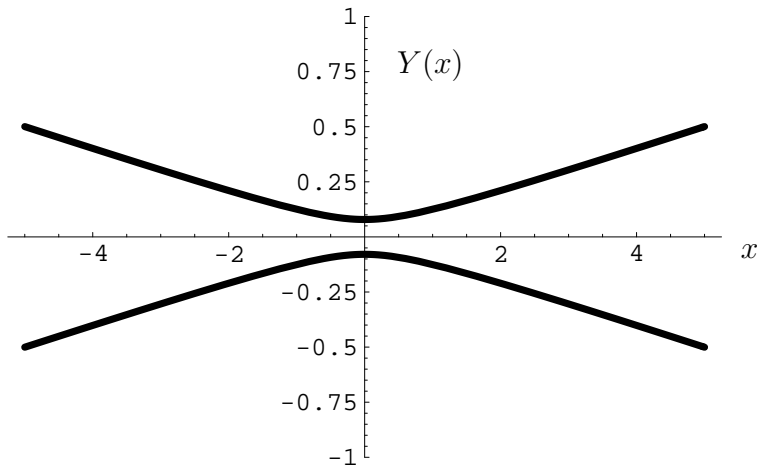}
\put(-130,150){$Y(x)$}
\put(0,80){$x$}
\caption{Intersecting D-strings are recombined. Here we draw the
 configuration
%(\ref{Ysol}) 
(2.18)
with parameters $q=\tan (\theta / 2)=0.1$
 and $C=\sqrt{q}/4$ (with $\pi \alpha' = 1 $).
}
\label{recombfig}
\end{center}
\end{minipage}
\end{center}
\end{figure}
%}

Keeping this condition in mind, let us study the geometric picture of
the condensation of the tachyonic fluctuation. The locations of the
D-strings are specified by the eigenvalues of the Higgs field $\varphi$. 
We diagonalize the whole Higgs field including the fluctuation
(\ref{gauss}), as\footnote{The gauge fields are not diagonalized
simultaneously.}
\begin{eqnarray}
\Phi (t,x)=\frac12 \left(
\begin{array}{cc}
qx & \varphi_0(x) \\ \varphi_0(x) & - qx
\end{array}\right) 
\rightarrow 
\frac12
\left(
\begin{array}{cc}
\sqrt{q^2 x^2 +C^2 e^{-qx^2}} & 0 \\ 
0 & -\sqrt{q^2x^2+C^2 e^{-qx^2}}
\end{array}\right) .
\end{eqnarray}
Therefore after the tachyon condensation, two D-strings are located
at 
\begin{eqnarray}
 Y(x)=\pm \pi\alpha' \sqrt{q^2x^2+C^2 e^{-qx^2}}
  = \pm \sqrt{
\left(\tan\left(\theta/2\right)x\right)^2 
+ C^2 \exp \left[-\frac{\tan(\theta/2)}{\pi\alpha'}x^2\right]
} \ \ \ \ .
\label{Ysol}
\end{eqnarray}
This configuration of D-strings is shown in Fig.~\ref{recombfig}. The
intersecting  D-strings are recombined! Naively it might have been
expected that the recombination might be difficult to be realized in Yang-Mills
theory since the locations of the D-strings are represented by the eigen
values and the recombination process must interchange two eigenvalues only
in a half of the worldvolume ($x<0$). However this is possible, 
as seen in the above diagonalization, due to the gauge transformation
which was singular at $x=0$ originally before the tachyon condensation 
(the limit $C \rightarrow 0$). So essentially the tachyon condensation
resolves the singularity of the gauge transformation and enables the
interchange of the left ``halves'' of the D-strings (the $x<0$ region).

The intriguing point here is that the magnitude of the tachyon mode $C$
corresponds just to the interbrane separation. This can be easily seen
in Eq.~(\ref{Ysol}) with $x=0$. Therefore, in the D-brane recombination,
the tachyon condensation is geometrically realized. 

\begin{figure}[tp]
\begin{center}
\begin{minipage}{12cm}
\begin{center}
\includegraphics[width=10cm]{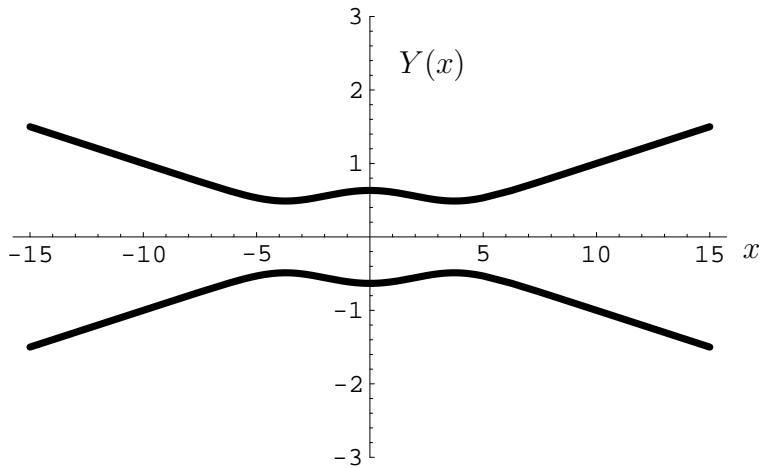}
\put(-130,150){$Y(x)$}
\put(0,80){$x$}
\caption{For large $C$, physically unacceptable recombination 
 is provided by (2.18). Here we have chosen the same value of $q$ 
 as in the previous figure but with $C=2\sqrt{q} > \sqrt{q}$. This
 exceeds the validity of the 
 approximation and thus prohibited. Before the configuration reaches
 this form, the nonlinear effect of fluctuations becomes
 important. }
\label{clargefig}
\end{center}
\end{minipage}
\end{center}
\end{figure}

It is interesting to note that the condition (\ref{constraint}) has
the following geometrical meaning. From the expression (\ref{Ysol})
of the transverse displacement, we observe that when the tachyon
develops the condensation and $C$ becomes large enough,  
the D-string worldvolume embedding is ``waving'' in the bulk (See
Fig.~\ref{clargefig}). This situation is quite strange in the naive view
of the 
energy minimization in which the D-string energy would be given mostly
as a tension times its length. The ``wave'' should be smeared.
The requirement that this ``wave'' does not appear is equivalent to the
condition that the function $Y(x)$ (\ref{Ysol}) has only a single
extremum. This condition is easily evaluated as $q>C^2$ which is
satisfied by (\ref{constraint}).

%%%%%%%%%%%%%%%%%%%%%%%%%%%%%%%%%%%%%%%%%%%%%%%%%%%%%%%%%%%%%
%%%%%%%%%%%%%%%%%%%%%%%%%%%%%%%%%%%%%%%%%%%%%%%%%%%%%%%%%%%%%
%%%%%%%%%%%%%%%%%%%%%%%%%%%%%%%%%%%%%%%%%%%%%%%%%%%%%%%%%%%%%
\setcounter{footnote}{0}
\section{Large intersection angle and Sen's conjecture}

In the previous section we showed that the recombination of the
intersecting D-strings is realized in Yang-Mills theory, 
the low energy scheme of string theory. However as mentioned in the
introduction, this phenomenon of recombination should be related to the
tachyon condensation of brane-antibrane system. The worldsheet picture
(\ref{1}) tells us that when $\theta=\pi$, the tachyon mass squared
reaches its lower bound $m^2 = -1/2\alpha'$ which is well-known as that
of tachyons whose condensation is conjectured to make the
brane-antibrane disappear (the Sen's conjecture, see
Ref.~\cite{Senconje}).  

\begin{figure}[bhtp]
\begin{center}
\begin{minipage}{13cm}
\begin{center}
\includegraphics[width=9cm]{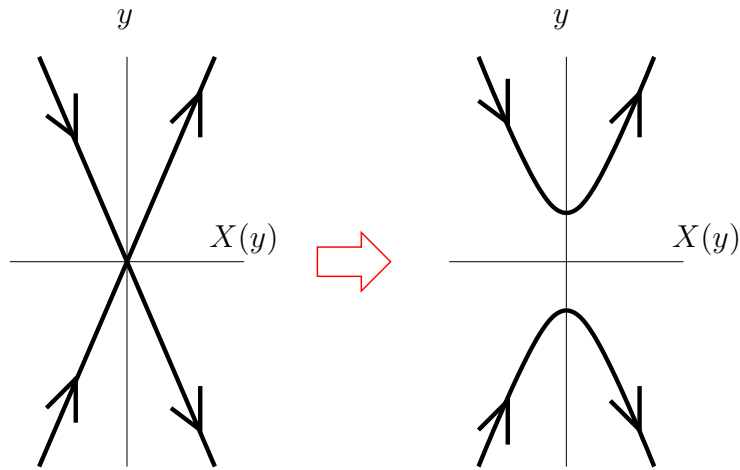}
\put(-50,170){$y$}
\put(-5,85){$X(y)$}
\put(-215,170){$y$}
\put(-180,85){$X(y)$}
\caption{D-string recombination with $\theta \sim \pi$. By rotating 
 these figures by $\pi/2$, one can see the brane-antibrane
 annihilation occurring around the origin.}
\label{tachyonfig}
\end{center}
\end{minipage}
\end{center}
\end{figure}

Then, how the disappearance of the brane-antibrane is related to the
recombination? The answer is found when we consider D-strings
intersecting at large $\theta$, $\theta \sim \pi$. For this large
$\theta$, the worldvolume parametrization with $x$ (the horizontal axis)
is not appropriate. Instead, we exchange the roles of $x$ and $Y(x)$ : 
now let us parametrize the worldvolume of D-strings by $y$, the vertical
axis.\footnote{This kind of the change of the axes of the
worldvolume parametrization has been investigated in Ref.~\cite{HHM} as
``target space rotation.''}
Then the displacement of the D-strings from this axis is measured
by a scalar field $X(y)$ which was originally $x$ (See the left 
of Fig.~\ref{tachyonfig}). One notices that in this parametrization, two
D-strings become a pair of a D-string and an {\it anti-D-string}, since
the 
orientation of one D-string is reversed compared to the other, as is
obvious in Fig.~\ref{tachyonfig}.

The tachyonic mode $n=0$ of (\ref{1}) remains at the origin and tends to
be condensed. Since we have explicitly shown in the previous section 
that the condensation of
this mode is equivalent to the D-string recombination, so assuming that 
even with large $\theta$ this equivalence is maintained, the
condensation of this tachyonic mode for large $\theta$ 
should provide again the
recombination depicted in the right of Fig.~\ref{tachyonfig}.
Reminding of the fact that the natural parametrization of the
worldvolume is $y$, we observe from this figure that around $y=0$
where the tachyon is localized the brane-antibrane has disappeared!
We regard this as a qualitative verification of the Sen's conjecture on
the tachyon condensation. 

The single assumption which we made in the argument above was that 
the identification of the fluctuation mode (\ref{gauss}) with the 
lowest mode of the worldsheet spectrum (\ref{1}) is valid also for large
$\theta$. Although it is hard to imagine that this is violated, in the
following we provide another evidence that the identification for 
large $\theta$ is valid. 

For larger $\theta$, the approximation adopted in the
previous section is getting worse, and so we cannot use the Yang-Mills
approach.  Hence we necessarily have to use alternative picture mentioned
above, the worldvolume parametrization in terms of $y$. 
This brane-antibrane picture is difficult to be analysed
since it needs stringy description with the energy scale $\alpha'$, such
as string field theories.
However, we shall see that a very simple tachyon action 
derived from string scattering amplitudes turns out to be useful for our
purpose. The tachyon two point function on a D$p$-antiD$p$ pair 
is reproduced by the free theory
\begin{eqnarray}
 S = -\int\! dtd^py
\left[\frac{1}{2}|\p_\mu T|^2-\frac{1}{4 \alpha'} |T|^2 
\right] \ .
\end{eqnarray}
Here the tachyon field is  complex since the strings joining the
D-strings are orientable. This tachyon is charged under a combination 
of the gauge fields, 
$A\equiv A^{(1)}-A^{(2)}$ where $A^{(i)}_\mu$ is the $U(1)$ 
gauge field on the $i$-th D$p$-brane ($i=1,2$). Taking this into account,
the above action is generalized to be gauge-invariant,
\begin{eqnarray}
S=-\int\! dtd^py \left[
\frac14 F_{\mu\nu}^2 + \frac{1}{2}|D_\mu T|^2-\frac{1}{4 \alpha'} |T|^2 
\right] \ ,
\end{eqnarray}
where the covariant derivative is defined as 
\begin{eqnarray}
 D_\mu T \equiv \p_\mu T - i A_\mu T \ .
\end{eqnarray}
For describing the nontrivial transverse displacement $X(y)$ from the
$y$ axis, we make a
T-duality (dimensional reduction) in which $X(y)$ is identified as a
gauge field, $2X(y)=2\pi\alpha'A_1$.
The resultant action is 
\begin{eqnarray} 
S=-\int\! dtdy \left[
\frac14 F_{\mu\nu}^2 
+\frac{2}{(2\pi\alpha')^2}(\p_\mu X)^2
+\frac{1}{2}|D_\mu T|^2
-\frac{1}{4 \alpha'} |T|^2 
+\frac{2}{(2\pi\alpha')^2} X^2|T|^2 \right] \ .
\label{T-du}
\end{eqnarray}
This action is defined on a 1+1 dimensional worldvolume spanned by $y$
and $t$. The field $X(y)$ describes the distance between the 
(anti-)D-string and the $y$ axis.
A solution which represents the intersecting D-strings 
is obtained as
\begin{eqnarray}
X(y)= \tan (\phi /2) y \ , \quad T=0 \ , \quad
F_{ab}=0 \ ,
\label{Tsol}
\end{eqnarray}
where we have defined $\phi=\pi-\theta$ which is small.

This solution (\ref{Tsol}) is expected to modify the tachyon mass
squared and give the localization of the tachyon around $y=0$. Let us
check these. The real part of the 
tachyon fluctuation $T$ around the solution
(\ref{Tsol}) is decoupled from other fluctuations, and the Lagrangian
for this fluctuation is given by\footnote{Since the solution is $T=0$,
we denote the tachyon fluctuation also as $T$.} 
\begin{eqnarray}
S_{\text{fluc}}=-\int\! d^2x 
\left[
\frac{1}{2} (\p_\mu T)^2 
-\frac{1}{4 \alpha'} T^2+
\frac{2 \tan^2 (\phi /2) y^2}{(2\pi\alpha')^2} T^2 
\right] .
\end{eqnarray}
The eigenvalue problem of this harmonic potential is easily solved 
to give the lowest mass squared which is negative, 
\begin{eqnarray}
m^2=-\frac{1}{2\alpha'}
+\frac{1}{\pi \alpha'}  \tan (\phi /2) 
= -\frac{\theta}{2\pi\alpha'}  +{\cal O}\left((\pi-\theta)^2\right)
\ .
\label{lowestT}
\end{eqnarray}
This agrees with the lowest tachyonic mode in the string spectrum
(\ref{1}).\footnote{The fluctuation analysis gives an equally-spaced
spectrum $m^2=-\frac{1}{2\alpha'}
+\frac{2k+1}{\pi \alpha'}  \tan (\phi /2)$ with a non-negative integer 
$k$. Note that this $k$ does not coincide with $n$ in
Eq.~(\ref{1}). Only the lowest mode $k=0$ corresponds to the 
lowest mode $n=0$.} 

The eigen function for this tachyonic mode (\ref{lowestT}) is found as 
\begin{eqnarray}
 T = \widetilde{C}(t) 
\exp\left[-\frac{\tan(\phi/2)}{2\pi\alpha'} y^2
\right]
\label{gaussT}
\end{eqnarray}
which is localized at $y=0$. Interestingly, this wave function of the
tachyonic mode is precisely the same as that of the lowest fluctuation
of the Yang-Mills analysis (\ref{gauss}). 
To show this, first we have to
note that the asymptotic configuration of the D-strings is 
given in both cases (\ref{Ysol}) and (\ref{Tsol}) by 
\begin{eqnarray}
 y = \pm \tan(\theta/2) x 
\label{iden1}
\end{eqnarray}
where we have used the definition $\phi \equiv \pi - \theta$ which gives 
\begin{eqnarray}
 \tan(\theta/2) = \left[\tan(\phi/2)\right]^{-1}.
\label{iden2}
\end{eqnarray}
This strongly suggests that the wave function of the tachyonic
fluctuation mode obtained in Yang-Mills scheme (\ref{gauss}) is not
deformed even for large $\theta$. In addition, the tachyon condensation
parameter $\widetilde{C}$ is expected to be proportional to $C$.

As we mentioned, other fluctuations such as the scalar fluctuation
$\delta X(y) \equiv X(y) - 2\tan(\phi/2)y$ 
is decoupled from the tachyon fluctuation. 
But once the tachyonic fluctuation is condensed 
as (\ref{gaussT}) with nonzero $\widetilde{C}$, this induces nontrivial
effect on $\delta X$. One can solve this backreaction to $X(y)$ provided
the tachyon condensation (\ref{gaussT}). This gives 
some information on the brane disappearance around $y=0$. 
The result is 
\begin{eqnarray}
 \delta X(y) 
\sim \frac{\pi\alpha'}{2}\widetilde{C}^2\! \int^y \! d\hat{y} \;
\exp\left[-\frac{\tan(\phi/2)}{\pi\alpha'} \hat{y}^2
\right]\ .
\label{xfluc}
\end{eqnarray}
Interestingly, this is an error function and thus the region where 
$\delta X$ is non-vanishing is not localized. Since the localized tachyon
condensation should not give nonlocal translation of the D-strings, 
the above result indicates that the worldvolume gets discontinuous
around $y=0$ : the result (\ref{xfluc}) should be split into two
regions so that the effect is localized, 
\begin{eqnarray}
 \delta X(y) \sim 
\left\{
\begin{array}{rl}
-\displaystyle
\frac{\pi\alpha'}{2}\widetilde{C}^2\! \displaystyle
\int^y_{-\infty} \! d\hat{y} \;
\exp\left[-\frac{\tan(\phi/2)}{\pi\alpha'} \hat{y}^2
\right] & (y<0)
 \\[15pt]
 \displaystyle\frac{\pi\alpha'}{2}\widetilde{C}^2\! \displaystyle
\int^\infty_{y} \! d\hat{y} \;
\exp\left[-\frac{\tan(\phi/2)}{\pi\alpha'} \hat{y}^2
\right] & (y>0)
\end{array}
\right. \ .
\label{deltax}
\end{eqnarray}
We have added different constant parameters for the two regions so that
asymptotically ($y \rightarrow \pm\infty$) $\delta X(y)$ vanishes.
Noting the following rough approximation for large $z$
\begin{eqnarray}
 \int^\infty_z \!d\hat{z}\; e^{- c \hat{z}^2} \sim 
\frac{1}{2c z}e^{-c z^2},
\end{eqnarray}
$\delta X(y)$ in Eq.~(\ref{deltax}) is approximated for large $y$ as
\begin{eqnarray}
 \delta X(y) \sim 
\left\{
\begin{array}{rl}
-\displaystyle
 \frac{(\pi\alpha')^2}{4\tan(\phi/2)}\widetilde{C}^2\cdot
\frac1{y}
\exp\left[-\frac{\tan(\phi/2)}{\pi\alpha'} y^2
\right] & (y<0)
 \\[15pt]
\displaystyle
 \frac{(\pi\alpha')^2}{4\tan(\phi/2)}\widetilde{C}^2\cdot
\frac1{y}
\exp\left[-\frac{\tan(\phi/2)}{\pi\alpha'} y^2
\right] & (y>0)
\end{array}
\right. \ .
\label{resultexp}
\end{eqnarray}
Surprisingly, this form coincides with the Yang-Mills result (\ref{Ysol})
for large $x$ as seen below. We expand (\ref{Ysol}) for large $x$ as 
\begin{eqnarray}
\frac{Y(x)}{\tan(\theta/2)} 
= \pm
\left(x \; + \; \frac{C^2}{2\tan^2(\theta/2)}
\frac1{x}\exp \left[-\frac{\tan(\theta/2)}{\pi\alpha'}x^2\right]
+ \mbox{higher}
\right) \ .
\label{largex}
\end{eqnarray}
Rewriting this with the identification used previously,
(\ref{iden1}) and (\ref{iden2}), one can show that the second term in 
Eq.~(\ref{largex}) is identical with (\ref{resultexp})
with a certain linear relation between 
the two constant parameters $C$ and $\widetilde{C}$. 
The second term in Eq.~(\ref{largex}) is of order $C^2$ due to the
diagonalization (\ref{Ysol}). To compare this with the
tachyon picture, we had to consider the backreaction which is higher
order in $\widetilde{C}$.

With the simple tachyon action (\ref{T-du}) employed, we have shown
various coincidence with the fluctuation analysis of the previous
section.  It would be interesting to study this tachyon system more
carefully, for example with use of the boundary string field theory
for brane-antibranes. 
A trial along this direction is presented in Appendix B.

%%%%%%%%%%%%%%%%%%%%%%%%%%%%%%%%%%%%%%%%%%%%%%%%%%%%%%%%%%%%%
%%%%%%%%%%%%%%%%%%%%%%%%%%%%%%%%%%%%%%%%%%%%%%%%%%%%%%%%%%%%%
%%%%%%%%%%%%%%%%%%%%%%%%%%%%%%%%%%%%%%%%%%%%%%%%%%%%%%%%%%%%%

\section{Dynamical aspects of recombination: decay width}

We have identified the tachyonic modes which triggers the D-brane
recombination and showed its significance in relation to the Sen's
conjectures. In application to D-brane cosmology for example, it is
important to study how the time-dependent recombination process
proceeds. This question is related to the recent development on rolling
tachyons \cite{rolling}, 
however our privilege is that we can use low energy
description instead of fully following the worldsheet string theory.
The recombination process itself is just the vacuum condensation of the
tachyonic mode (\ref{gauss}), however once the tachyon expectation value
$C(t)$ reaches its critical value $\sqrt{q}$, the nonlinear effect (or
equivalently, higher order terms of fluctuations) becomes important.
After a while, the off-diagonal modes of the stretched open string will
decouple and the system will be dictated by a Nambu-Goto action
for a D-string. The relative velocity between D-strings leaving each
other will soon reach the speed of light (in our approximation of
decoupling limit of closed strings), as discussed in Ref.~\cite{KH}. 
This is a geometrical realization of rolling tachyon \cite{rolling},
since in the present case the magnitude of the tachyon mode is identical
with the interbrane separation, as seen in Section 2.

In the argument above we started from intersecting D-strings, however
this configuration is of course unstable and it is another question
whether this is a reasonable initial condition to be started with or
not. In the context of string cosmology, it is expected that first many
D-branes are created in a hot universe and then they interact with
each other by recombination. Therefore more physical initial condition
might be non-parallelly separated D-branes. 

\begin{figure}[tp]
\begin{center}
\includegraphics[width=10cm]{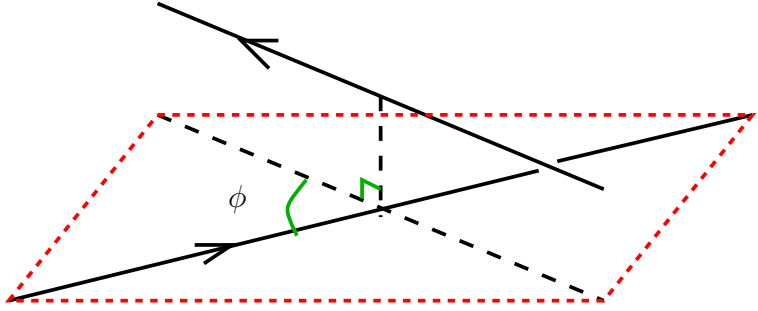}
\put(-200,37){$\phi$}
\caption{Intersecting branes are separated. }
\label{separatefig}
\end{center}
\end{figure}

In this section 
we present a naive estimation of the decay width of the non-parallelly
separated D-branes. 
In fact the following process might be physically plausible: two 
D-strings which are not parallel approach each other by gravitational
and Ramond-Ramond forces, and just at the time when they intersect, the tachyon
field at the intersection point starts rolling down the potential hill
and condenses to provide the recombination. However, as discussed in
Ref.~\cite{KH, tye}, before they intersect there should be a small
probability for quantum tunnelling directly into the recombined final
state. 
For analysing this process, 
we place two D-strings apart from each other so that the
open string connecting the two does not yet have any
tachyonic excitation. We fix the distance between the branes as in
Fig.~\ref{separatefig} for simplicity. 
In this situation, the branes cannot approach each other and
thus the decay mode is due to the quantum tunnelling effect.
We generalize the computation in Ref.~\cite{KH} to
the case of the branes not only separated anti-parallelly 
but also tilted against each other. The intersection angle 
$\phi$ will appear in the decay width. 

The quantum tunnelling is dictated by a bounce solution whose geometric
interpretation is the Euclidean worldvolume deformation of the D-string,
that is, a formation of a throat connecting the two D-strings.
When $\phi=0$, the D-strings are placed anti-parallelly, and the decay
width can be evaluated in the following way:
First, the decay width is the exponential of the on-shell Euclidean 
action, 
\begin{eqnarray}
 \Gamma = \exp (-S_{\rm E})
\end{eqnarray}
where $S_{\rm E}$ is the action into which the throat (volcano) 
solution of the worldvolume field theory is substituted (see
Ref.~\cite{CM} for this type of solutions). 
The throat action is evaluated easily as
\begin{eqnarray}
 S_{\rm E}= T_{\rm D1}\left(2\pi R d - 2 \pi R^2\right) \ ,
\end{eqnarray}
where $d$ is the distance between the branes, and $R$ is the radius of
the throat. The first term is from the area of the cylinder (throat), 
and the second term is from the areas of the cap and the bottom of the 
throat which are holes and thus give a negative contribution.
The throat radius $R$ can be determined by extremization of the
action  $S_{\rm E}$ as
\begin{eqnarray}
 \frac{\p S_{\rm E}}{\p R} = 0 \ ,
\end{eqnarray}
which gives $R=d/2$. Then the on-shell action is
\begin{eqnarray}
 S_{\rm E} = T_{\rm D1}\frac{\pi d^2}{2} \ .
\end{eqnarray}
Thus a trivial result is that when the distance between the two becomes
large, the decay width by a formation of a throat is exponentially
dumping, as $\Gamma \sim \exp(-T_{\rm D1} d^{p+1})$. 

Now let us introduce the intersection angle to the system. 
It is expected that the location of the throat formed by the tunnel
effect is unique, at the intersection point. One way to explain this is
that, from the geometrical point of view, it is unlikely for the throat
to be formed far from the intersection point since it needs extra energy 
to create it (in other words, the potential barrier is higher). Another
way to explain this is that the string connecting the two is localized
around the intersection point, as seen from both the Yang-Mills picture 
and the tachyon picture studied in the previous sections.

So here we simply assume that the throat is formed at the intersection
point (though they are not actually intersecting but separated by $d$).
Further assumption is that the cylinder is squeezed and become
elliptic due to the localization of the string. We assume that the
horizontal section of this cylinder which was a circle for $\phi=0$
is deformed to be an ellipse,
\begin{eqnarray}
\tau^2 +  \frac{y^2}{f^2} = R^2 \ .
\label{elip}
\end{eqnarray}
Here $\tau$ is the Euclidean time, and 
$f$ is a function of $\phi$ which will be determined later
by the extremization process.

The area of two holes can be easily evaluated as 
\begin{eqnarray}
 S_{\rm E}^{\rm (hole)} = -T_{\rm D1}2\pi R^2 f
\end{eqnarray}
where the last factor $f$ comes from the elliptic deformation. 
On the other hand, the area of the side of the cylinder is
\begin{eqnarray}
 S_{\rm E}^{\rm (side)} = T_{\rm D1}\cdot
 4\int_0^R d\tau  \sqrt{1 + \left(\frac{dy}{d\tau}\right)^2}
\sqrt{d^2 + 4\tau^2\tan^2 (\phi/2)} \ ,
\end{eqnarray}
where $y$ is now defined by $\tau$ through (\ref{elip}).
The equations of motion for $f$ and $R$ are given by
\begin{eqnarray}
 \frac{\p}{\p R} S_{\rm E} = 0, 
\quad  \frac{\p}{\p f} S_{\rm E} = 0
\label{ex}
\end{eqnarray}
where $S_{\rm E} =  S_{\rm E}^{\rm (side)} +S_{\rm E}^{\rm (hole)}$.
These are integral equations, 
and the integrands are difficult to be exactly evaluated. 
So let us concentrate on only small $\phi$. 
It is natural to expect a correction
\begin{eqnarray}
 f = 1 +  {\cal O} (\phi^2) >1
\end{eqnarray}
which means that the cylinder is squeezed, in other words, enlarged
along the direction $y$. So we define $1 + 2g \equiv f^2$ and expand the
integrand so that the integration is possible. The result for the first
equation in (\ref{ex}) is 
\begin{eqnarray}
&& 2\pi d
\left[
1 + \frac12\left(g + 6\frac{R^2}{d^2}\tan^2(\phi/2)\right)
\right.
\nonumber\\
&&
\hspace{10mm}
\left.
+ \frac38
\left(
-\frac12 g^2\! +\! 6g\frac{R^2}{d^2}\tan^2(\phi/2)
\!-\!10 \frac{R^4}{d^4}\tan^4(\phi/2)
\right)
-\frac{2R}{d}\left(
1\!+\!g\!-\!\frac12 g^2
\right)
\right] =0 \ .\nonumber
\end{eqnarray}
The second equation in (\ref{ex}) gives
\begin{eqnarray}
 \frac{\pi}{8}dR\!
\left[
8\! +\! 2g\! +\! 12 \frac{R^2}{d^2}\tan^2(\phi/2)
\!-\!\frac52 g \!+\! 2 g \frac{R^2}{d^2}\tan^2(\phi/2)
\!-\!10 \frac{R^4}{d^4}\tan^4(\phi/2) \!-\!16\frac{R}{d}
\right]\!=\!0 \ .
\nonumber
\end{eqnarray}
In both equations we neglect ${\cal O} (\phi^6)$ terms.
Then these two can be solved as 
\begin{eqnarray}
&& R = \frac{d}{2}\left(
1+ \frac12 \tan^2(\phi/2)+ \frac38 \tan^4(\phi/2)
 + {\cal O}(\phi^6)
\right) \ ,
\label{Rr}
\\
&& g = \frac{1}{2}\tan^2(\phi/2) + \frac12 \tan^4(\phi/2) 
+ {\cal   O}(\phi^6) \ .
\end{eqnarray}
Hence the cylinder is squeezed as expected ($g>0$),
\begin{eqnarray}
  f = 1 + \frac12 \tan^2(\phi/2) + \frac38 \tan^4(\phi/2)
+ {\cal   O}(\phi^6) \ .
\label{fr}
\end{eqnarray}
The final on-shell action into which these values are substituted is
\begin{eqnarray}
  S_{\rm E}  = T_{\rm D1} \frac{\pi d^2}{2}
\left(1 + \frac{1}{2}\tan^2(\phi/2) + \frac38 \tan^4(\phi/2)
+ {\cal O}(\phi^6)\right).
\label{Sr}
\end{eqnarray}
This result shows that a larger $\phi$ gives a smaller 
probability of the throat formation.\footnote{
It is interesting to note that all the above results (\ref{Rr}),
(\ref{fr}) and (\ref{Sr}) has the same corrections in $\phi$. 
This indicates some deeper physical reason. We may expect that the
corrections can be summarized universally as
\begin{eqnarray}
\left(1 + \frac{1}{2}\tan^2(\phi/2) + \frac38 \tan^4(\phi/2)
+ {\cal O}(\phi^6)\right) = F(\phi)
\end{eqnarray}
with 
$ \lim_{\phi \rightarrow \pi} F(\phi) = +\infty$,
so that the system approaches BPS parallel D-branes giving a vanishing
decay width. 
} This result is consistent with the observation that,
a system closer to a supersymmetric configuration of parallel
parallel two D-strings ($\phi=\pi$) is more stable so the most
unstable system should be the 
parallel D1-antiD1 configuration ($\phi=0$). 
As $\phi$ increases, the
tachyonic instability of the system decreases and thus the throat
nucleation probability should decrease accordingly.

We note here that we cannot
compare the above result with the case with 
exactly vanishing $\phi$, since $\phi=0$ case gives a
probability per a unit world volume of D-strings while in our present
case the formation occurs only at the intersection point. So the
physical meaning of the resultant decay width is different from the
$\phi=0$ case.

The width we have found here is very small, since 
we are working in the decoupling limit $g_{\rm string} \rightarrow 0$. 
This is potentially due to the fact that the distance between the
D-strings is fixed. One of the other possibilities of the initial
conditions is to consider
D-strings approaching each other with a constant relative velocity $v$.
If this $v$  is small enough, there should be enough long period for the
D-strings to be recombined by
developing the tachyon condensation. The period during which 
the tachyonic mode is present is evaluated as 
$\sim \theta l_{\rm string}/v$. The condensation should start by
a quantum fluctuation studied in Ref.~\cite{GuthPi}, whose
characteristic time is roughly $\sim l_{\rm string}/\theta$,
hence we expect that the D-strings recombination occurs 
when $\theta^2>v$.

%%%%%%%%%%%%%%%%%%%%%%%%%%%%%%%%%%%%%%%%%%%%%%%%%%%%%%%%%%%%%
%%%%%%%%%%%%%%%%%%%%%%%%%%%%%%%%%%%%%%%%%%%%%%%%%%%%%%%%%%%%%
%%%%%%%%%%%%%%%%%%%%%%%%%%%%%%%%%%%%%%%%%%%%%%%%%%%%%%%%%%%%%

\section{Future directions}

The mechanism of the recombination of the intersecting D-branes provided
in this paper is quite simple, basically due to the fact that we can
employ low energy approximation of the D-brane dynamics which is the
super Yang-Mills theory. Therefore we expect that it is possible to apply 
the recombination mechanism presented in this paper
in many directions which include
(i) study for formulating string/M-theory in terms of higher dimensional
D-branes as fundamental objects, (ii) verification of the Sen's
conjectures using various schemes, and (iii) string phenomenology based
on intersecting branes, and brane cosmology.

As for (i), the first step might be to generalize the present analysis
to higher dimensional D-branes.\footnote{Instabilities in the higher
dimensional cases were discussed in Ref.~\cite{Uranga} in the context of
M-theory lift. } 
For example, D2-branes intersecting in 
$T^4$ can be supersymmetric with a certain relation of two
intersection angles \cite{BDL}. It is interesting to see this phenomena
in terms of low energy Yang-Mills theory.  

Regarding (ii), since there has been vast amount of literature concerning
the Sen's conjectures, the relation between those and
our mechanism should be explored further. 
We may turn on the intersection angle
for example in the scheme of cubic string field theories or boundary
string field theories, which will help to clarify the relation. On the
other hand, our analysis shows that the tachyon condensation of
brane-antibrane can be seen by the Yang-Mills theories, 
which suggests that they can describe various
phenomena associated with the tachyon condensation, such as the brane 
descent relations and creation of lower-dimensional D-branes. 
Furthermore, it might be possible even to describe all the tachyon 
condensation processes by (generalizations of) Yang-Mills theories.
For example, non-BPS branes defined as unstable defects in tachyon
condensation 
of brane-antibrane can also be described in our Yang-Mills scheme,
and this might be the case also for the rolling tachyons \cite{rolling}
and associated 
S(pacelike)-branes \cite{sbrane} 
whose deformations describe creation of lower
dimensional defects \cite{saction}. 
In our realization, the rolling tachyon is
realized geometrically as recombined D-branes leaving each
other. The rolling tachyon in the present case should be described by 
a Dirac-Born-Infeld action in which the scalar field is identified with
the tachyon,  as seen in Section 2. This might give a path to justify
the tachyon action of Ref.~\cite{rolling}. 

Finally, there should be much possibilities for application to
phenomenology and cosmology (iii) since nowadays intersecting branes are
considered to be an effective way of realizing our world on branes.
Recombination of topological defects has been largely studied especially 
in the context of cosmic string network formation in cosmology
\cite{book}. In 
string theory, BPS D-branes can be realized as topological solitons in  
unstable D-branes with tachyons, based on the Sen's conjectures
\cite{Senconje} 
on tachyon condensation. This viewpoint realizes 
the formation and time evolution of D-brane networks in the early
universe, and the clarification of the recombination mechanism in this
paper will help analysing phenomenological parameters. 
We hope that our research may be the first step to mediate between 
the string phenomenology and the tachyon condensation in string theory.

%%%%%%%%%%%%%%%%%%%%%%%%%%%%%%%%%%%%%%%%%%%%%%%%%%%%%%%%%%%%%
%%%%%%%%%%%%%%%%%%%%%%%%%%%%%%%%%%%%%%%%%%%%%%%%%%%%%%%%%%%%%
%%%%%%%%%%%%%%%%%%%%%%%%%%%%%%%%%%%%%%%%%%%%%%%%%%%%%%%%%%%%%

\acknowledgments{
We would like to thank Washington Taylor IV for quite valuable discussions
which were indispensable for completion of the present work,
and appreciate useful comments by Akikazu Hashimoto, Tatsuya Noguchi 
and Gary Shiu.
K.~H. is supported in part by the Grant-in-Aid for Scientific
Research (No.~12440060 and 13135205) from the Japan Ministry of
Education, Science and Culture.
After this paper was published, we were informed of the paper
\cite{Morosov} in which instability of intersecting branes was studied 
in Yang-Mills theory.}

%%%%%%%%%%%%%%%%%%%%%%%%%%%%%%%%%%%%%%%%%%%%%%
%%%%%%%%%%%%%%%%%%%%%%%%%%%%%%%%%%%%%%%%%%%%%%
%%%%%%%%%%%%%%%%%%%%%%%%%%%%%%%%%%%%%%%%%%%%%%

\appendix

\section{Higher order in $\theta$ and NBI corrections }

Here we discuss how the higher order terms in $\theta$ in the
string theory spectrum (\ref{1}) 
can be reproduced from the non-Abelian Born-Infeld action. 
The first nontrivial corrections to the Yang-Mills theory 
appear as $F^4$ terms in a
symmetrized trace \cite{tse}, 
\begin{eqnarray}
 L = {\rm Str} 
\sqrt{-\det(\eta_{\mu\nu} + 2\pi\alpha' F_{\mu\nu})} \ .
\end{eqnarray}
The Lagrangian with the transverse scalar fields can be obtained simply
by T-dualities (dimensional reductions). 
To compute the higher order corrections, it is of great help to use the
expansion of Ref.~\cite{AkiWati} which provided a general
expression of the fluctuation Lagrangian expanded around a diagonal and
constant field strength background in $U(2)$ case. Here we can make use
of it.  The final form of the fluctuation Lagrangian in
Ref.~\cite{AkiWati} is  
\begin{eqnarray}
 L = {\rm Str} 
\sqrt{-\det(\eta_{\mu\nu} + 2\pi\alpha' (F_0)_{\mu\nu})}
\left[\frac12 B^{\mu\nu}\tilde{F}_{\nu\mu}
-\frac14 g^{\mu\nu} \tilde{F}_{\nu\lambda} g^{\lambda \sigma}
\tilde{F}_{\sigma\mu}
\right]
\end{eqnarray}
where $F_0$ is the background value of the field strength, 
$g$ is the open string metric, and 
$B^{\mu\nu}\equiv -g^{\mu\lambda} F_{\lambda\rho}\eta^{\rho\nu}$.
In our case, the worldvolume is 1+2 dimension before taking the
T-duality down to 1+1 dimensions, 
so we turn on only the field strength $(F_0)^{12}$.
Then the open string metric reads
\begin{eqnarray}
 g^{\mu\nu} = {\rm diag}(-1,g,g) \ , \quad
g\equiv \frac{1}{1+(2\pi\alpha' (F_0)_{12})^2} \ .
\end{eqnarray}
Then the fluctuation Lagrangian is written as
\begin{eqnarray}
 L &= &{\rm Str}
\frac{1}{\sqrt{g}}
\biggl(
g\cdot (F_0)_{12} i[ A_1, \varphi])
\biggr.
\nonumber \\
& &\hspace{10mm}
\left.
-\frac12 g(\p_0 A_1^a)^2 -\frac12 g(\p_0 \varphi^a)^2 
-\frac12 g^2 (\p_1 \varphi^a-i[A_1,\varphi]^a)^2
\right) \ .
\label{lines}
\end{eqnarray}
We note here that the first line in the above expression receives no
combinatoric factor when the symmetrized trace is taken, but the second
line receives it. The prescription is given in Ref.~\cite{AkiWati}.
When one has a symmetrized trace over a function of the form 
$(F_0)^m \tilde{F}(F_0)^n\tilde{F}$, then the $F_0$ dependence should
appear as 
\begin{eqnarray}
 f(F_0) \; \rightarrow \; \frac{1}{F_0} \int_0^{F_0}\! dx\; f(x) \ .
\end{eqnarray}
Following this prescription, we have 
\begin{eqnarray}
&& \sqrt{g} \rightarrow
1-\frac16 (2\pi\alpha'(F_0)_{12})^2 
-\frac3{40}
(2\pi\alpha'(F_0)_{12})^4 
+ {\cal O}\left((2\pi\alpha'(F_0)_{12})^6\right)\ ,
\\
&& g^{3/2} \rightarrow
1-\frac12 (2\pi\alpha'(F_0)_{12})^2 
-\frac3{40}
(2\pi\alpha'(F_0)_{12})^4 
+ {\cal O}\left((2\pi\alpha'(F_0)_{12})^6\right) \ ,
\end{eqnarray}
while the first line in Eq.~(\ref{lines}) gives an expression as it is,
\begin{eqnarray}
 \sqrt{g} F_0 = F_0 
\left[1-\frac12 (2\pi\alpha'(F_0)_{12})^2 
-\frac3{8}
(2\pi\alpha'(F_0)_{12})^4 
+ {\cal O}\left((2\pi\alpha'(F_0)_{12})^6\right)
\right].
\end{eqnarray}
Finally we obtain the fluctuation Lagrangian as 
(in the following we write only the terms up to the order $F^4$
explicitly, since it is known that the symmetrized trace prescription
for corrections is valid only at this order \cite{AkiWati})
\begin{eqnarray}
 L &=& \left( 1-\frac16 (2\pi\alpha'(F_0)_{12})^2 \right)
\left[
 -\frac12 (\p_0A)^2 -\frac12 (\p_0\varphi)^2
\right.
\nonumber
\\
& &\hspace{20mm}\left.
-\frac12
\left( 1-\frac13 (2\pi\alpha'(F_0)_{12})^2 \right)
\left(
(\p_1\varphi)^2 + 2qxA\p_1\varphi + q^2 x^2 A^2 -2qA\varphi
\right)
\right]
\nonumber
\\
& &
+ {\cal O}\left((2\pi\alpha'(F_0)_{12})^4\right) \ .
\label{finally}
\end{eqnarray}
So, up to the overall factor which is irrelevant for the fluctuation
analysis, the difference between the present corrected Lagrangian and
(\ref{fluctu}) is only 
that the mass squared will be multiplied by a factor
appearing in the second line of Eq.~(\ref{finally}). 
Noting that the correspondence here is
that $(F_0)_{12}= q/2$,
our result of the fluctuation analysis is 
\begin{eqnarray}
m^2 = (2n-1)q\left( 1-\frac13 (2\pi\alpha'q)^2 \right) 
= \left(n-\frac12\right)\frac{\theta}{\pi\alpha'} 
+ {\cal O} (\theta^5) \ .
\end{eqnarray}
This coincides with the worldsheet result (\ref{1}), up to 
${\cal O} (\theta^5)$ ambiguity.

If we assume that the action is given by the symmetrized trace at 
order $F^6$, we can soon find that the mass spectrum is incorrect as was  
shown in Ref.~\cite{AkiWati}. The non-Abelian BI action at order $F^6$
was proposed in some articles \cite{F^6action} and 
we can use our fluctuation analysis for the first nontrivial check
of their $F^6$ terms \cite{Nagaoka}.

%%%%%%%%%%%%%%%%%%%%%%%%%%%%%%%%%%%%%%%%%%%%%%%%%%%%%%%%%%
%%%%%%%%%%%%%%%%%%%%%%%%%%%%%%%%%%%%%%%%%%%%%%%%%%%%%%%%%%
%%%%%%%%%%%%%%%%%%%%%%%%%%%%%%%%%%%%%%%%%%%%%%%%%%%%%%%%%%

\section{Tachyon mass squared to higher order in $\phi$}

In this appendix, we present a trial to obtain higher order corrections
to the tachyon mass eigen values obtained in Section 3. To proceed, we
somehow assume the form of the higher order coupling in the
tachyon-gauge system of the brane-antibrane. 
A trial is the following action with the higher correction of gauge
fields, 
\begin{eqnarray}
S=- \int d^3x \left(-\frac{1}{4 \alpha'} |T|^2 +\frac{1}{2} 
G^{ab} D_a T D_b \bar{T}
\right) \ ,
\label{genet}
\end{eqnarray}
where $G^{ab}$ is the open string metric
\begin{eqnarray}
G^{ab}=\left(\frac{1}{\eta+2\pi\alpha'F}\right)_{\text{sym}} \ .
\end{eqnarray}
This fluctuation action can be obtained from 
the action for example,
\begin{eqnarray}
S=-2T \int dx^3
V(|T|)\sqrt{-\det (g_{ab}+F_{ab}+D_{\{a} TD_{b\}} \bar{T})} \ .
\end{eqnarray}
This action is the generalization of a non-BPS D-brane action to 
the brane-antibrane system. The antisymmetric part of 
$D_a T D_b \bar{T}$ is dropped by the reality condition.

Now, let us check the order $O(\phi^2)$ of the mass spectrum
in this action (\ref{genet}).
We proceed in the same manner as in Section 3, 
to obtain the fluctuation Lagrangian
\begin{eqnarray}
S=-\int dtdy
\Big(
-\frac{1}{4 \alpha'} T^2 +\frac{1}{2} (\p_\mu T)^2 
+\displaystyle\frac{[(\p_y T)^2+4 \tan^2 (\phi /2) y^2 T^2]}
{2(1+4 \tan^2 (\phi /2))}
\Big) \ .
\end{eqnarray}
{}From this the mass squared is obtained as 
\begin{eqnarray}
m^2=\frac{\tan (\phi /2)}{\pi \alpha' (1+4 \tan^2 (\phi /2))}(2n+1)-
\frac{1}{2 \alpha'} \ ,
\end{eqnarray}
and for the lowest mode, it is given by
\begin{eqnarray}
\sim \frac{-\theta}{2 \pi \alpha'} +O \left((\pi-\theta)^3\right) \ .
\end{eqnarray}
Therefore we have obtained the correct mass spectrum up to 
$O\left((\pi-\theta)^3\right)$ ambiguity.

%%%%%%%%%% References %%%%%%%%%%%%%%%%%%%%%%%%%
\newcommand{\J}[4]{{\sl #1} {\bf #2} (#3) #4}
\newcommand{\andJ}[3]{{\bf #1} (#2) #3}
\newcommand{\AP}{Ann.\ Phys.\ (N.Y.)}
\newcommand{\MPL}{Mod.\ Phys.\ Lett.}
\newcommand{\NP}{Nucl.\ Phys.}
\newcommand{\PL}{Phys.\ Lett.}
\newcommand{\PR}{ Phys.\ Rev.}
\newcommand{\PRL}{Phys.\ Rev.\ Lett.}
\newcommand{\PTP}{Prog.\ Theor.\ Phys.}
\newcommand{\hep}[1]{{\tt hep-th/{#1}}}
%%%%%%%%%%%%%%%%%%%%%%%%%%%%%%%%%%%%%%%%%%%%%%%

\end{document}